\documentclass[12pt]{iopart}
\usepackage{graphicx,setstack}

\begin{document}

\title[Eternal acceleration and negative $\Lambda$]{Halting
eternal acceleration with an effective negative cosmological constant}

\author{V.F. Cardone\footnote{Corresponding author\,: {\tt winnyenodrac@gmail.com}.}}
\address{Dipartimento di Fisica "E.R. Caianiello",
Universit\'{a} di Salerno, via S. Allende, 84081\,-\,Baronissi (Salerno),
Italy}

\author{R.P. Cardenas, Y. Leyva Nodal}
\address{Departamento de Fisica, Universidad Central de Las
Villas, Santa Clara, CP, 54830\,-\,Villa Clara, Cuba}

\begin{abstract}
In order to solve the problem of eternal acceleration, a model has been
recently proposed including both a negative cosmological constant $\Lambda$
and a scalar field evolving under the action of an exponential potential.
We further explore this model by contrasting it against the Hubble diagram
of Type Ia supernovae, the gas mass fraction in galaxy clusters and the
acoustic peak and shift parameters. It turns out that the model is able to
fit quite well this large dataset so that we conclude that a negative
$\Lambda$ is indeed allowed and could represent a viable mechanism to halt
eternal acceleration. In order to avoid problems with theoretical
motivations for both a negative $\Lambda$ term and the scalar field, we
reconstruct the gravity Lagrangian $f(R)$ of a fourth order theory of
gravity predicting the same dynamics (scale factor and Hubble parameter) as
the starting model. We thus end up with a $f(R)$ theory able to both fit
the data and solve the problem of eternal acceleration without the need of
unusual negative $\Lambda$ and ad hoc scalar fields.
\end{abstract}


\section{Introduction}

Recent astrophysical data, from the Hubble diagram of Type Ia Supernovae
(SNeIa) \cite{SNeIa,Riess04} to the measurement of the anisotropy and
polarization spectra of the cosmic microwave background radiation (CMBR)
\cite{wmap,VSA} and large scale structure data \cite{LSS}, point out
towards a picture of the universe unexpected only few years ago. According
to this new scenario, the universe is spatially flat, undergoing a phase of
accelerated expansion and dominated by a smoothly distributed negative
pressure fluid dubbed {\it dark energy}. Although a lot of candidates has
been proposed for this mysterious component \cite{QuintRev}, the
concordance $\Lambda$CDM model \cite{Lambda} made out of cold dark matter
(CDM) and the Einstein cosmological constant $\Lambda$ stands as the best
fit to a wide range of different astrophysical data
\cite{Teg03,Sel04,Teg06}. Notwithstanding its observational merits, the
$\Lambda$CDM scenario is seriously plagued by the well known coincidence
and fine tuning problems which are the main motivations to look for
alternative models.

Although as a classical (general relativistic) field theory the
concordance model is fairly simple, a universe presently dominated
by a positive $\Lambda$ term is on the contrary quite difficult to
understand from a quantum theory viewpoint. Indeed, since matter
and radiation energy density fades away as $a^{-3}$ and $a^{-4}$
respectively (with $a$ the scale factor), the universe turns out
to be asymptotically dominated by the vacuum energy accounted for
by the positive $\Lambda$. As a consequence, the universe
asymptotically enters a de Sitter phase with the scale factor
growing exponentially and a never ending accelerated expansion.
Let us then consider the definition of cosmic horizon as\,:

\begin{equation}
\delta \propto \int_{t_0}^{t_{end}}{\frac{cdt}{a(t)}} < \infty
\label{horizon}
\end{equation}
with $t_0$ and $t_{end}$ the present day age of the universe and its age at
the end of time (i.e., $a(t_{end} = 0$). Note that, while for a closed
universe $t_{end}$ takes a finite value, this is not for open and spatially
flat models ($t_{end} \rightarrow \infty$). For a de Sitter like universe,
$\delta$ takes a finite value so that a cosmic horizon appears. It is worth
stressing that this problem is not strictly related to $\Lambda$, but turns
out to be a consequence of eternal acceleration. Should we replace
$\Lambda$ with a quintessence scalar field, the universe should still be
eternally accelerated finally reaching a de Sitter phase and hence again a
finite cosmic horizon.

The presence of a cosmic horizon implies that it is not possible to define
pure state vectors of quantum asymptotic states. As a consequence, in a de
Sitter background spacetime, it is not possible to define a gauge invariant
scattering matrix ${\cal{S}}$. First studied in \cite{sasaki}, this problem
is known as {\it superexpansionary divergence} in the context of quantum
field theory (QFT) and is particularly troublesome for the formulation of
strings theory. Indeed, by construction \cite{strings}, perturbative string
theory is based on the well defined nature of scattering amplitudes of
various excitations and hence fundamentally relies on the possibility to
correctly define the ${\cal{S}}$\,-\,matrix \cite{smatrix}. As such,
formulating string theory in presence of a finite cosmic horizon is
challenging. The so called Liouville string framework \cite{ddk} represents
an attempt to solve the problem in the case of non conformal backgrounds
(including the de Sitter spacetime), but much work is still needed before a
complete mathematical formulation is achieved.

A radically different approach is, however, possible. Indeed, observations
do tell us that the universe is spatially flat and accelerating, but this
by no way implies that the acceleration should be eternal. This latter
feature turns out only as a consequence of assuming a positive $\Lambda$
term\footnote{A similar argument also holds for a quintessence scalar field
whatever is the self interaction potential.} to explain cosmic speed up.
From an observational point of view, however, nothing prevents us to
consider an effective dark energy fluid as a source of the accelerated
expansion. We can then split its energy density as the sum of two
components and then investigate the consequences on the future evolution of
the universe. Such an approach has been recently investigated by some
authors \cite{Gu,RolLambda,Grande} with interesting results. In particular,
in \cite{RolLambda}, some of us have presented a model whose dark energy
component is the sum of a {\it negative} cosmological constant and a
quintessence scalar field evolving under the action of an exponential
potential. As a result, although the model is presently accelerating,
eternal acceleration disappears and the universe ends in a Big Crunch like
singularity in a finite time. Motivated by these theoretical virtues, we
further explore this model by the observational point of view in order to
see whether a negative $\Lambda$ is indeed compatible with the
astrophysical data at hand.

Notwithstanding the positive evidence of the model, it is also worth
wondering whether an alternative theoretical derivation is possible.
Actually, there are difficulties to reconcile a negative $\Lambda$ term
within the framework of Quantum Field Theory. In the context of the
Standard Model of Elementary Particle Physics, Spontaneous Symmetry
Breaking induces a large (negative) value of the cosmological constant:

\begin{equation}
\Lambda_{ind} = -\frac{M_H^2}{8\sqrt{2}G_F}
\end{equation}
If we apply the current experimental bounds on the mass of the Higgs boson
$M_H$ and on the Fermi constant $G_F$, we obtain a (negative) value 55
orders of magnitude larger than the observed upper bound of the
cosmological constant CC. There have been several attempts to solve this
problem, without enough success \cite{Weinberg}. A vacuum cosmological term
$\Lambda_v$ with (positive) opposite sign is often introduced, so that the
physical observable cosmological constant results\,:

\begin{equation}
\Lambda_{ph} = \Lambda_{ind} + \Lambda_v
\end{equation}
The introduction of $\Lambda_v$ is also dictated by the requirement of
renormalizability of the massive theory, so the CC problem is rather the
need of the extremely precise choice of the corresponding normalization
condition $\Lambda_{ph} = 0$ in the very far infrared. It is very difficult
to explain why the two terms on the right hand side of above equation
should cancel each other with such accuracy at that point. On the other
hand, as for all quintessence models, it is not totally clear where the
scalar field comes from. Having such a large dominance by dark ingredients
is, for sure, a disturbing situation so that it is worth to look for other
potential explanations.

Following the prescriptions of the Occam razor, it is desirable to have as
less components of unknown origin as possible in a cosmological model. In
order to overcome this problem, we resort to fourth order theories of
gravity (also referred to in the following as $f(R)$ theories). According
to this approach, cosmic acceleration is the first signal of a breakdown of
General Relativity so that the gravity Lagrangian has to be modified by
replacing the scalar curvature $R$ with a generic analytic function $f(R)$
\cite{capozcurv,review,noicurv,MetricRn}. Moreover, it has been shown
\cite{CCT} that it is always possible to find out a $f(R)$ theory
reproducing the same dynamics (i.e., scale factor and Hubble parameter) as
a given dark energy model. Using this method, we can reformulate the model
above in terms of fourth order theories of gravity. The resulting $f(R)$
will allow to both fit the data and solve the problem of eternal
acceleration without resorting to a controversial negative $\Lambda$ and an
ad hoc scalar field.

The plan of the paper is as follows. We present the starting cosmological
model in Sect. 2 where we generalize the approach in \cite{RolLambda}
including the radiation term. The numerical solution for the expansion rate
for this model is the key ingredient for the likelihood analysis discussed
in Sect. 3 where we contrast the model against a wide set of observational
data thus being able to constrain its characteristic parameters. Although
our model comprises two ingredients other than standard matter and
radiation, it is worth stressing that it is dynamically equivalent to a
model dominated by a single dark energy fluid whose equation of state is
reconstructed in Sect. 4. The best fit model is, on the other hand, used as
input for the procedure of reconstructing $f(R)$ which is presented in
Sect. 5 where we show that a negative $\Lambda$ could result as part of an
effective representation of a fourth order theory. Sect. 6 is finally
devoted to conclusions.

\section{The model}

Looking at the impressive amount of papers addressing the problem of cosmic
acceleration clearly shows that two leading candidates to the dark energy
throne are the old cosmological consntat $\Lambda$ and a scalar field
$\phi$ evolving under the influence of its self\,-\,interaction potential
$V(\phi)$. While a $\Lambda$ term adds a constant energy density
$\rho_{\Lambda}$ and a negative pressure $p_{\Lambda} = -\rho_{Lambda}$
into the cosmic evolution equations, a scalar field partecipates to the
dynamics through its energy density and pressure given as\,:

\begin{equation}
\left \{
\begin{array}{lll}
\rho_{\phi} & = & \displaystyle{\frac{1}{2} \dot{\phi}^2 + V(\phi)} \\
~ & ~ & ~ \\ p_{\phi} & = & \displaystyle{\frac{1}{2} \dot{\phi}^2 -
V(\phi)} \\
\end{array}
\right .
\label{eq: rphi}
\end{equation}
with a dot denoting the derivative with respect to the cosmic time $t$. The
evolution of the scalar field is then governed by its Klein\,-\,Gordon
equation\,:

\begin{equation}
\ddot{\phi} + 3 H \dot{\phi} + \frac{dV}{d\phi} = 0
\label{eq: kg}
\end{equation}
with $H = \dot{a}/{a}$ the Hubble parameter and $a$ the scale factor
(normalized to unity at the present time). It is easy to show that
Eq.(\ref{eq: kg}) is the same as the continuity equation for the scalar
field energy density $\rho_{\phi}$ under the hypothesis that $\phi$ does
not interact with the matter and the other ingredients of the cosmic pie.

In the usual approach, one adds either a scalar field or a cosmological
constant term to the field equations. However, since what we see is only
the final effect of the dark energy components, in principle nothing
prevents us to add more than one single component provided that the
effective dark energy fluid coming out is able to explain the data at hand.
Moreover, as we have hinted upon above, a single scalar field, while
explaining cosmic speed up, leads to a problematic eternal acceleration. A
possible way out of this problem has been proposed by some of us
\cite{RolLambda} through the introduction of a negative cosmological term.
Motivated by those encouraging results, we therefore consider a spatially
flat universe filled by dust matter, radiation, scalar field and a
(negative) cosmological constant term. The Friedmann equations thus read\,:

\begin{equation}
H^2 = \frac{1}{3} \left [ \rho_M + \rho_r + \rho_{\Lambda} +
\frac{1}{2} \dot{\phi}^2 + V(\phi) \right ] \ ,
\label{eq: f1}
\end{equation}

\begin{equation}
2 \dot{H} + 3 H^2 = - \left [ \frac{1}{3} \rho_r - \rho_{\Lambda} +
\frac{1}{2} \dot{\phi}^2 - V(\phi) \right ] \ ,
\label{eq: f2}
\end{equation}
where we have used natural units with $8 \pi G = c = 1$. Using the
continuity equation for $\rho_M$ and $\rho_r$ and the definition of density
parameter for the i\,-\,th fluid

\begin{displaymath}
\Omega_i = \frac{\rho_i(a = 1)}{\rho_c} = \frac{\rho_i(a = 1)}{3H_0^2} \ ,
\end{displaymath}
Eqs.(\ref{eq: f1}) and (\ref{eq: f2}) may be rewritten as\,:

\begin{equation}
E^2 = \Omega_M a^{-3} + \Omega_r a^{-4} + \frac{1}{2 H_0^2} \left ( \frac{d
\tilde{\phi}}{dt} \right )^2 + \tilde{V}(\tilde{\phi}) \ ,
\label{eq: f1bis}
\end{equation}

\begin{equation}
H_0^{-1} \dot{E} = - \frac{3}{2} \Omega_M a^{-3} - 2 \Omega_r a^{-4} -
\frac{1}{2 H_0^2} \left ( \frac{d\tilde{\phi}}{dt} \right )^2 \,
\label{eq: f2bis}
\end{equation}
with $E = H/H_0$, $\tilde{\phi} = \phi/\sqrt{3}$, $\tilde{V} = V/3H_0^2$,
and we denote with a subscript $0$ the present day value of a quantity. It
is convenient to introduce the dimensionless variable

\begin{displaymath}
u = \ln{(1 + z)} = - \ln{a}
\end{displaymath}
with $z = 1/a - 1$ the redshift (having set $a_0 = 1$). In terms of this
variable, the Friedmann equations become\,:

\begin{equation}
E^2 = \frac{\Omega_M {\rm e}^{3u} + \Omega_r {\rm e}^{4u} +
\Omega_{\Lambda} + \tilde{V}(\tilde{\phi})}{1 - (1/2) \left ( d\tilde{\phi}/du \right )^2} \ ,
\label{eq: f1u}
\end{equation}

\begin{equation}
\frac{1}{2} \frac{dE^2}{du} = \frac{3}{2} \Omega_M {\rm e}^{3u} + 2 \Omega_r {\rm e}^{4u}
+ \frac{3}{2} E^2 \left ( \frac{d\tilde{\phi}}{du} \right )^2 \ ,
\label{eq: f2u}
\end{equation}
while the Klein\,-\,Gordon equation reads\,:

\begin{equation}
E^2 \frac{d^2\tilde{\phi}}{du^2} + \left ( \frac{1}{2} \frac{dE^2}{du} - 3
E^2 \right ) \frac{d\tilde{\phi}}{du} + \frac{d\tilde{V}}{d\tilde{\phi}} =
0 \ .
\label{eq: kgu}
\end{equation}
As well known, given an expression for the scalar field potential
$\tilde{V}(\tilde{\phi})$, only two out of the three equations (\ref{eq:
f1u}), (\ref{eq: f2u}), (\ref{eq: kgu}) are independent. It is convenient
to insert Eqs.(\ref{eq: f1u}) and (\ref{eq: f2u}) into Eq.(\ref{eq: kgu})
to get a single equation governing the evolution of the scalar field\,:

\begin{equation}
E^2 \frac{d^2\tilde{\phi}}{du^2} + \left [ \frac{3}{2} \Omega_M {\rm
e}^{3u} + \Omega_r {\rm e}^{4u} + 3 \Omega_{\Lambda} + 3
\tilde{V}(\tilde{\phi}) \right ] \frac{d\tilde{\phi}}{du} = 0
\label{eq: sfeq}
\end{equation}
with $E^2(u)$ given by Eq.(\ref{eq: f1u}). In order to solve (numerically)
this equation thus determining $\tilde{\phi}(u)$ and then E(u) through
Eq.(\ref{eq: f1u}), one has to set two initial conditions. The first one
can be trivially obtained by evaluating Eq.(\ref{eq: f1u}) in $u = 0$ to
get\,:

\begin{displaymath}
\tilde{V}_0 = \Omega_{\phi} - \frac{1}{2} \left ( \frac{d\tilde{\phi}}{du}
\right )^2_{u = 0}
\end{displaymath}
with $\tilde{V}_0 = \tilde{V}(\tilde{\phi}_0)$. Given the shape of the
potential $V(\phi)$, the above relation can be inverted to get the present
day value of the scalar field. A second initial condition can be obtained
by first considering the  equation of state (hereafter EoS) of the field
given as\,:

\begin{equation}
w_{\phi} = \frac{p_{\phi}}{\rho_{\phi}} = \frac{E^2 \left (
d\tilde{\phi}/du \right )^2 - 2 \tilde{V}}{E^2 \left ( d\tilde{\phi}/du
\right )^2 + 2 \tilde{V}} \ .
\label{eq: wphi}
\end{equation}
Evaluating this at the present day and using the above relation for
$\tilde{\phi_0}$, we finally get the initial conditions\,:

\begin{equation}
\tilde{V}_0 = \frac{\Omega_{\phi}}{2} (1 + w_0) \ ,
\label{eq: phiz}
\end{equation}

\begin{equation}
\left ( \frac{d\tilde{\phi}}{du} \right )^2_{u = 0} = \Omega_{\phi} (1 - w_0)
\label{eq: dphiduz}
\end{equation}
with $w_0 = w_{\phi}(u = 0)$. Note that, since the left hand side of
Eqs.(\ref{eq: phiz}) and (\ref{eq: dphiduz}) are positive definite, we get
the constraint $-1 \le w_0 \le 1$ which is always verified for any ordinary
scalar field whatever the potential $\tilde{V}(\tilde{\phi})$ is.

\subsection{The deceleration parameter}

In order to separate models in accelerating and decelerating ones, one has
to compute the present day value of the deceleration parameter defined
as\,:

\begin{displaymath}
q(u) = - \frac{\ddot{a} a}{\dot{a}^2} = - 1 - \frac{\dot{H}}{H^2} = -1 +
\frac{1}{E} \frac{dE}{du} \ .
\end{displaymath}
Using Eqs.(\ref{eq: f1u}) and (\ref{eq: f2u}), we can therefore write the
deceleration parameter as\,:

\begin{eqnarray}
q = -1 & + & \left [ 1 - \left ( \frac{d\tilde{\phi}}{du} \right )^2 \right
] \\ ~ & {\times} & \frac{(3/2) \Omega_M {\rm e}^{3u} + 2 \Omega_r {\rm e}^{4u} +
(3/2) \left ( d\tilde{\phi}/du \right )^2}{\Omega_M {\rm e}^{3u} + \Omega_r
{\rm e}^{4u} + \Omega_{\Lambda} + \tilde{V}(\tilde{\phi})} \ . \nonumber
\label{eq: qu}
\end{eqnarray}
Evaluating this relation in $u = 0$ and using Eq.(\ref{eq: dphiduz}), the
present day value then reads\,:

\begin{equation}
q_0 = \frac{1}{2} (1 + \Omega_r) - \frac{3}{2} (\Omega_{\Lambda} +
\Omega_{\phi} w_0) \ ,
\label{eq: qz}
\end{equation}
so that, in order to have accelerating models $(q_0 \le 0)$, we have to set
the constraint\,:

\begin{equation}
w_0 \le \frac{\Omega_{\Lambda} - (1 + \Omega_r)/2}{\Omega_{\phi}} \ .
\label{eq: qzconst}
\end{equation}
We can also solve Eq.(\ref{eq: qz}) with respect to $w_0$ to get\,:

\begin{equation}
w_0 = - \frac{(1 - 2q_0) + \Omega_r - 3 \Omega_{\Lambda}}{3 \Omega_{\phi}}
\end{equation}
so that the initial conditions now rewrite\,:

\begin{equation}
\tilde{V}_0 = (3\Omega_{\phi} + \Omega_r - 3 \Omega_{\Lambda} + 1 - 2 q_0)/6 \ ,
\end{equation}

\begin{equation}
\left ( \frac{d\tilde{\phi}}{du} \right )^2_{u = 0} = 3 \Omega_{\phi} - \Omega_r +
3 \Omega_{\Lambda} - 1 + 2 q_0 \ .
\end{equation}
Since the lhs of these relations are positive quantities, we get the
following constraints on the present day deceleration parameter\,:

\begin{equation}
\frac{1}{2} (1 + \Omega_r) + \frac{3}{2} (\Omega_{\phi} + \Omega_{\Lambda}) \le q_0 \le
\frac{1}{2} (1 + \Omega_r) + \frac{3}{2} (\Omega_{\phi} - \Omega_{\Lambda}) \ .
\label{eq: qzrange}
\end{equation}
For accelerating models, the lower limit in Eq.(\ref{eq: qzrange}) must be
negative so that we get\,:

\begin{equation}
\Omega_{\Lambda} \le \frac{1}{3} (1 + \Omega_r) - \Omega_{\phi} \ .
\label{eq: olconst}
\end{equation}
Note that for models with a negative cosmological constant this relation is
always satisfied thus meaning that for such a choice it is possible to work
out accelerating solutions of the field equations.

\subsection{The exponential potential}

The above discussion holds whatever is the scalar field potential
$V(\phi)$, but in order to actually solve the cosmic equations we have to
definitively assign an analytical expression for this quantity. Following
\cite{RS,RolExp}, we consider an exponential potential\,:

\begin{equation}
V(\phi) = B^2 \exp{(-\sigma \phi)}
\label{eq: phipot}
\end{equation}
with $B^2$ a generic constant and $\sigma^2 = 12 \pi G/c^2 = 3/2$. As shown
in \cite{RStest}, such a choice leads to a cosmological model (made out of
the above scalar field and dust matter) in good agreement with a large set
of astrophysical data.

Although being interesting on both theoretical and observational grounds,
such a model is however affected by the problem of eternal acceleration. As
can be easily understood, since the matter energy density decreases as
$a^{-3}$ (with $a$ the scale factor normalized to be unity at the present
day), the universe becomes soon scalar field dominated and the accelerating
expansion never ends. In order to avoid this problem, Cardenas et al.
\cite{RolLambda} have added a third ingredient to the cosmic pie, namely a
negative cosmological constant. The N\"{o}ther symmetry approach \cite{Noether}
makes it possible to find a convenient change of variables in such a way
that exact solutions are found.

Although dealing with an analytical expression is welcome, the result in
\cite{RolLambda} only holds if the radiation term is neglected. While this
is not a problem when considering the late universe, introducing radiation
drastically changes the structure of the cosmic equations so that the
N\"{o}ther symmetry approach does not apply anymore. As a consequence, the
analytical solution may no more be used and a numerical analysis is needed.
To this end, we insert Eq.(\ref{eq: phipot}) into Eq.(\ref{eq: phiz}) and
solve with respect to $\tilde{\phi}_0$\,:

\begin{equation}
\tilde{\phi}_0 = - \sqrt{\frac{2}{9}} \ln{\left [ \frac{3 H_0^2}{2 B^2}
\Omega_{\phi} (1 - w_0) \right ]} = - \sqrt{\frac{2}{9}} \ln{\left [
\Omega_{\phi} (1 - w_0) \right ]}
\label{eq: phizexp}
\end{equation}
where, in the second equality, we have arbitrarily set $B^2 = (3/2) H_0^2$.
Note that such a choice has no effect on the dynamics since it is a simple
rescaling of the scalar field $\tilde{\phi}$ which does not influence any
physically interesting quantity. With such a choice, the potential and its
derivative then read\,:

\begin{equation}
\left \{
\begin{array}{lll}
\tilde{V}(\tilde{\phi}) & = & \displaystyle{\frac{1}{2}
\exp{\left ( - \frac{3}{\sqrt{2}} \tilde{\phi} \right )}} \\
~ & ~ & ~ \\ \displaystyle{\frac{d\tilde{V}}{d\tilde{\phi}}} & = &
\displaystyle{\frac{3}{2 \sqrt{2}}
\exp{\left ( - \frac{3}{\sqrt{2}} \tilde{\phi} \right )}} \ .
\end{array}
\right .
\label{eq: phidphiexp}
\end{equation}
Summarising, in order to determine the cosmic dynamics (i.e., the scale
factor and the Hubble parameter) for such a model with both a negative
cosmological constant $\Lambda$ and a scalar field $\phi$ with an
exponential potential added to the standard dust matter and radiation
terms, we have to numerically integrate Eq.(\ref{eq: sfeq}) with the
potential given by Eq.(\ref{eq: phidphiexp}) and the initial conditions
(\ref{eq: phizexp}) and (\ref{eq: dphiduz}). The solution for
$\tilde{\phi}(u)$ thus obtained may then be inserted into the potential and
then into Eq.(\ref{eq: f1u}) to determine the Hubble parameter which can be
further integrated to get the scale factor. In order to perform such a
scheme, one has to set the values of up to four parameters, namely the dust
and radiation density parameters $\Omega_M$ and $\Omega_r$, and the present
day values $(\Omega_{\phi}, w_0)$ of the scalar field energu density and
EoS respectively. In particular, one could use Eq.(\ref{eq: qzconst}) to
choose $w_0$ in such a way that the resulting model will be today
accelerating.

\section{Matching with the data}

Notwithstanding how well motivated it is, a whatever model must be able to
reproduce what is observed. This is particularly true for the model we are
considering because the presence of a negative cosmological constant
introduces a positive pressure term potentially inhibiting the cosmic speed
up. Moreover, contrasting the model against the data offers also the
possibility to constrain its characteristic parameters and estimate other
derived interesting quantities, such as $q_0$, the transition redshift
$z_T$ and the age of the universe $t_0$. Motivated by these considerations,
we will therefore fit our model to the dataset described below
parametrizing the model itself with the matter density $\Omega_M$, the
scalar field quantities $(\Omega_{\phi}, w_0)$ and the dimensionless Hubble
constant $h$ (i.e., $H_0$ in units of $100 \ {\rm km/s/Mpc}$), while we
will set the radiation density parameter as $\Omega_r = 10^{-4.3}$ as in
\cite{FP04} from a median of different values reported in literature.

\subsection{The method and the data}

In order to constrain the model parameters, we maximize the following
likelihood\,:

\begin{equation}
{\cal{L}} \propto \exp{\left [ - \frac{\chi^2({\bf p})}{2} \right
]} \label{eq: deflike}
\end{equation}
where ${\bf p} = (\Omega_M, \Omega_{\phi}, w_0, h)$ denotes the set of
model parameters and the pseudo\,-\,$\chi^2$ merit function reads\,:

\begin{eqnarray}
\chi^2({\bf p}) & = & \sum_{i = 1}^{N}{\left [ \frac{\mu^{th}(z_i, {\bf p}) - \mu_i^{obs}}
{\sigma_i} \right ]^2} + \sum_{i = 1}^{N}{\left [
\frac{f_{gas}^{th}(z_i, {\bf p}) - f_{gas,i}^{obs}}{\sigma_i} \right ]^2}
\nonumber \\ ~ & + &
\displaystyle{\left [ \frac{{\cal{A}}({\bf p}) - 0.474}{0.017} \right ]^2} +
\displaystyle{\left [ \frac{{\cal{R}}({\bf p}) - 1.70}{0.03} \right
]^2}  + \left ( \frac{h - 0.72}{0.08} \right )^2 \ .
\label{eq: defchi}\
\end{eqnarray}
Let us discuss briefly the different terms entering Eq.(\ref{eq:
defchi}). In the first one, we consider the distance modulus $\mu
= m - M$, i.e. the difference between the apparent and absolute
magnitude of an object at redshift $z$. This is given as\,:

\begin{equation}
\mu(z) = m - M = 25 + 5 \log{D_L(z)}
\label{eq: defmu}
\end{equation}
with $D_L(z)$ the luminosity distance (in Mpc) defined as\,:

\begin{equation}
D_L(z) = \frac{c}{H_0} (1 + z) \int_{0}^{z}{\frac{dz'}{E(z',{\bf
p})}} \ . \label{eq: dl}
\end{equation}
As input data, we use the SNeIa sample assembled in \cite{D07} by putting
on a common scale the data recently released from the SNLS \cite{SNLS} and
ESSENCE collaborations \cite{ESSENCE} and the higher redshift SNeIa
observed with HST in the GOODS survey \cite{Riess06}. As well known, the
SNeIa Hubble diagram is unable to determine the Hubble constant $H_0$ since
this quantity is degenerate with the (unknown) absolute magnitude $M$. As
such, Davis et al. have put all the SNeIa on the same distance scale using
$h = 0.656$. In our analysis, however, we will leave $h$ as a free
parameter so that Eq.(\ref{eq: defmu}) must be rewritten as\,:

\begin{equation}
\mu(z) = 25 + 5 \log{\left ( \frac{c}{H_0^{fid}} \right )} + 5 \log{\left ( \frac{h^{fid}}{h} \right )}
+ 5 \log{d_L(z)}
\label{eq: mufit}
\end{equation}
with $H_0^{fid} = 100 h^{fid} = 65.6 \ {\rm km/s/Mpc}$ and $d_L =
D_L/(c/H_0)$ the Hubble free luminosity distance. Note that the value of
$h$ determined by this fit will not be fully reliable because of the
degeneracy hinted above. However, such a problem will not affect the
estimates of the other parameters since we will marginalize over $h$ in the
analysis of the results.

The second term in Eq.(\ref{eq: defchi}) relies on the gas mass fraction in
galaxy clusters. We briefly outline here the method referring the
interested reader to the literature for further details
\cite{fgasbib,fgasapp,fgasdata}. Both theoretical arguments and numerical
simulations predict that the baryonic mass fraction in the largest relaxed
galaxy clusters should be invariant with the redshift (see, e.g.,
Ref.\,\cite{ENF98}). However, this will only appear to be the case when the
reference cosmology in making the baryonic mass fraction measurements
matches the true underlying cosmology. From the observational point of
view, it is worth noting that the baryonic content in galaxy clusters is
dominated by the hot X\,-\,ray emitting intra\,-\,cluster gas so that what
is actually measured is the gas mass fraction $f_{gas}$ and it is this
quantity that should be invariant with the redshift within the caveat
quoted above. Moreover, it is expected that the baryonic fraction in
clusters equals the universal ratio $\Omega_b/\Omega_M$ so that $f_{gas}$
should indeed be given by $b {\times} \Omega_b/\Omega_M$ where the multiplicative
factor $b$ is motivated by simulations that suggest that the gas fraction
is slightly lower than the universal ratio because of processes that
convert part of the gas into stars or eject it outside the cluster.

Following Ref.\,\cite{fgasdata07}, we adopt the concordance $\Lambda$CDM
model (i.e., with $\Omega_M = 0.3$, $\Omega_{\Lambda} = 0.7$, $h = 0.7$) as
reference cosmology in making the measurements so that the theoretical
expectation for the apparent variation of $f_{gas}$ with the redshift is
\cite{fgasdata07}\,:

\begin{equation}
f_{gas}(z) = \frac{K \gamma A(z) b(z)}{1 + s(z)} \frac{\Omega_b}{\Omega_M}
\left [ \frac{D_A^{\Lambda CDM}(z)}{D_A(z, {\bf p})} \right ]^{3/2} \ .
\label{eq: fgas}
\end{equation}
Some words are needed to explain the meaning of the different terms
entering the above equations. First, the two functions $b(z)$ and $s(z)$
take into account variations of the ratio $\Omega_b/\Omega_M$ due to
gastrophysics and star formation respectively. In a first approximation,.
they are described as linear functions of the redshift. However, in order
to not increase the number of parameters, we will set them to constant
values taking $b(z) = b_0 = 0.83$ and $s(z) = s_0 = 0.16 (h/0.70)^{1/2}$ in
agreement with \cite{fgasdata07}. While $K$ and $\gamma$ are normalizing
factors correcting for measurement related problems and can be put to the
constant values 1.0 and 1.05 respectively, $A(z)$ actually plays a more
important role. To understantd its origin, it is worth remembering that
$f_{gas}$ is typically measured at a given fraction of the cluster virial
radius. However, on the sky distances are measured in angular rather than
physical units with the conversion depending on the assumed cosmological
model. Since the reference model is different from the actual one, a
correction term must be included to account for the change in the radius.
Considering the scaling of the different quantities involved in the
measurement process, it is possible to show that the correction term may be
approximated as \cite{fgasdata07}\,:

\begin{equation}
A(z) = \left [ \frac{E(z) D_A(z, {\bf p})}{E^{\Lambda CDM}(z) D_A^{\Lambda
CDM}(z)} \right ]^{\eta}
\label{eq: defaz}
\end{equation}
with $E^{\Lambda CDM}(z) = \left [ \Omega_M (1 + z)^3 +
\Omega_{\Lambda} \right ]^{1/2}$ and $\eta = 0.214$. Note that, actually, because
of the small $\eta$ value, this term does not play a significant role in
the fitting process, while a key ingredient is the last one in Eq.(\ref{eq:
fgas}) given by the ratio of the angular diameter distances $D_A(z) =
D_L(z)/(1 + z)^2$ between the $\Lambda$CDM and the model to be tested. It
is worth noting that $f_{gas}(z)$ depends not only on the integrated Hubble
parameter, but also explicitly on the baryon and total matter density
parameters $\Omega_b$ and $\Omega_M$. In particular, baryogenesis
calculations contrasted to the observed abundances of primordial elements
puts a severe constraint on the physical baryon density $\omega_b =
\Omega_b h^2$. Using this method, Kirkman et al. \cite{Kirk} have
determined\,:

\begin{displaymath}
\omega_b = 0.0214 {\pm} 0.0020 \ .
\end{displaymath}
A rigorous analysis should be done letting $\omega_b$ as a free parameter
eventually including the above estimates as a prior. However, in order to
not increase the number of parameters to be determined, we will set
$\omega_b = 0.0214$ neglecting the small error.

The third term in the definition of $\chi^2$ takes into account the
measurement of the {\it baryonic acoustic oscillation} (BAO) peak in the
large scale correlation function at $100 \ h^{-1} \ {\rm Mpc}$ separation
detected by Eisenstein et al. \cite{Eis05} using a sample of 46748 luminous
red galaxies (LRG) selected from the SDSS Main Sample \cite{SDSSMain}.
Actually, rather than the position of acoustic peak itself, a closely
related quantity is better constrained from these data, namely the acoustic
peak parameter defined as \cite{Eis05}\,:

\begin{equation}
{\cal{A}} = \frac{\sqrt{\Omega_M}}{z_{LRG}} \left [
\frac{z_{LRG}}{E(z_{LRG})} y^2(z_{LRG}) \right ]^{1/3} \label{eq:
defapar}
\end{equation}
with $z_{LRG} = 0.35$ the effective redshift of the LRG sample,
and we have introduced the dimensionless coordinate distance
$y(z)$ defined as\,:

\begin{equation}
y(z) = \int_{0}^{z}{\frac{dz'}{E(z', {\bf p})}} \ .
\label{eq: defy}
\end{equation}
As it is clear, the ${\cal{A}}$ parameter depends not only on the
dimensionless coordinate distance (and thus on the integrated
expansion rate), but also on $\Omega_M$ and $E(z)$ explicitly
which removes some of the degeneracies intrinsic in distance
fitting methods. Therefore, it is particularly interesting to
include ${\cal{A}}$ as a further constraint on the model
parameters using its measured value \cite{Eis05}\,:

\begin{displaymath}
{\cal{A}} = 0.469 \left ( \frac{n_s}{0.98} \right )^{-0.35} {\pm} 0.017
\end{displaymath}
with $n_s$ the spectral index of the primordial density perturbations. For
$n_s = 0.95$ as determined by the WMAP 3rd year analysis \cite{WMAP3}, we
get ${\cal{A}} = 0.474 {\pm} 0.017$ as set in Eq.(\ref{eq: defchi}). A caveat
is in order here. As discussed in \cite{Eis05}, the measurement of the
position of the BAO peak from the correlation function relies somewhat on
having assumed the $\Lambda$CDM model to convert angular distances in
physical distances and in the computation of the reference (smoothed) power
spectrum. As a consequence, one should use with caution the above value for
${\cal{A}}$ when using a different cosmological model and rather directly
fitting the measured correlation function. However, it is expected that the
position of the BAO peak does not change too much in a different model
since it is measured at a relatively low redshift. Since a full analysis
will require the solution of the perturbation equations for our model
(which is outside our aims here), we will follow the common practice in
literature neglecting this problem and directly using the ${\cal{A}}$
parameter as an observational constraint included in the $\chi^2$ merit
function.

Let us consider the fourth term in Eq.(\ref{eq: defchi}) which relies on
the {\it shift parameter} \cite{BET97}\,:

\begin{equation}
{\cal{R}}({\bf p}) = \sqrt{\Omega_M} y(z_{LS}, {\bf p})
\label{eq: defr}
\end{equation}
with $z_{LS}$ the redshift of the last scattering surface which we compute
using the approximation given in \cite{HS96}. Using the WMAP3 data, Wang \&
Mukherjee \cite{WM06} have determined ${\cal{R}} = 1.70 {\pm} 0.03$ in very
good agreement with what is expected for the concordance $\Lambda$CDM
model. Note that such a result may argue in favour of models with a
negligible dark energy component at high redshift, but we defer this
discussion to the later analysis of the results.

Finally, the term depending on $h$ in Eq.(\ref{eq: defchi}) is only a
Gaussian prior on this quantity obtained by considering the model
independent estimate of the Hubble constant recovered by the HST Key
project \cite{Freedman}. This collaboration have measured $H_0$ using a
wide set of different local distance calibrators thus ending up with a
value which is claimed to fully take into account any possible systematic
error inherent to the peculiarities of each single method. Comfortably,
such a measurement turns out to be in good agreement with other (less
precise) methods relying on different physics and distance scales as the
time delay in lensed quasars \cite{H0lens} and the Sunyaev\,-\,Zel'dovich
effect \cite{H0SZ} in galaxy clusters.

In order to maximize the likelihood function ${\cal{L}}({\bf p})$, we
should compute it over a very fine grid in the four dimensional parameter
space $(\Omega_M, \Omega_{\phi}, w_0, h)$ and then interpolate the results
for values falling in between two grid points. This is quite time consuming
for a very fine grid so that we resort to a Monte Carlo Markov Chain (MCMC)
method running three chains with 30000 points each and assessing the
convergence using the Gelman\,-\,Rubin test (with $|R - 1| = 0.1$). It is
worth stressing that, in order to be sure that the MCMC runs into the
region with negative $\Lambda$, it is better to reparametrize the model
using $\Omega_{\Lambda}$ instead of $\Omega_{\phi}$ as model parameter
using the obvious relation\,:

\begin{displaymath}
\Omega_{\phi} = 1 - \Omega_M - \Omega_{r} - \Omega_{\Lambda} \ .
\end{displaymath}
Therefore, in the following, we will use $(\Omega_M, \Omega_{\Lambda}, w_0,
h)$ as the parameter space to be explored by the MCMC code. After cutting
the burn\,-\,in period, the final coadded chain contains $\simeq 88000$
points thus guaranteeing an efficient coverage of the interesting region of
the parameter space allowing us to compute the marginalized likelihood
functions for each parameter $p_i$\,:

\begin{equation}
{\cal{L}}_{p_i}(p_i) \propto \int{dp_1 \ldots \int{dp_{i - 1}
\int{dp_{i + 1} ... \int{dp_n {\cal{L}}({\bf p})}}}} \label{eq:
defmarglike}
\end{equation}
which is then normalized at unity at maximum. Under the Bayesian framework,
the best estimated for the parameter $p_i$ is given by the median of the
marginalized likelihood, while the $68$ and $95\%$ confidence ranges are
given as $(x_{1 \sigma}, y_{1 \sigma})$ and $(x_{2 \sigma}, y_{2 \sigma})$
with $x_{i \sigma}$ and $y_{i \sigma}$ computed by solving respectively the
equations\,:

\begin{displaymath}
\int_{p_{i, min}}^{x_{i \sigma}}{{\cal{L}}_{p_i}(p_i) dp_i} = \delta_{i}
\int_{p_{i, min}}^{p_{i, max}}{{\cal{L}}_{p_i}(p_i) dp_i} \ ,
\end{displaymath}

\begin{displaymath}
\int_{y_{i \sigma}}^{p_{i, max}}{{\cal{L}}_{p_i}(p_i) dp_i} = \delta_{i}
\int_{p_{i, min}}^{p_{i, max}}{{\cal{L}}_{p_i}(p_i) dp_i} \ ,
\end{displaymath}
with $\delta_{i} = (1 - 0.68)/2$ for $i = 1$ and $(1 - 0.95)/2$ for $i = 2$
and $(p_{i, min}, p_{i, max})$ the lower and upper bounds chosen for the
parameter $p_i$. For the model we are considering, we conservatively let
$\Omega_M$ range between 0.15 and 0.45, while the range for $h$ is $(0.45,
0.85)$. While the upper bound for $\Omega_{\Lambda}$ is dictated by our
constraint $\Lambda < 0$, choosing an upper limit is a more complicated
issue. Indeed, the scalar field and the negative cosmological constant may
be incorporated in a single dark energy fluid (as we will see later) so
that there is a degeneracy in balancing the two individual components. We
therefore arbitrarily set $\Omega_{\Lambda} > -1.5$ thus allowing values of
$\Omega_{\phi}$ quite larger than 1. Finally, we cut the physically
acceptable range $(-1, 1)$ for $w_0$ to become $(-1, 0)$ which is still a
conservative choice. Indeed, we could also use Eq.(\ref{eq: qzconst}) to
set an upper bound for $w_0$ ensuring $q_0 \le 0$. However, we prefer to be
fully open minded not forcing the chains to explore a priori acccelerating
models only.

\subsection{Results}

\begin{figure}
\centering \resizebox{12cm}{!}{\includegraphics{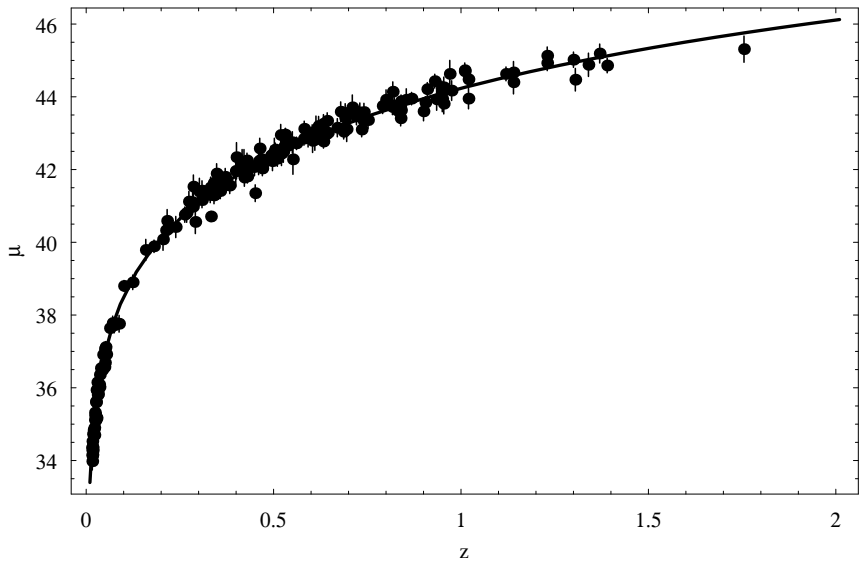}}
\caption{Best fit curve superimposed to the data on the SNeIa
Hubble diagram.}
\label{fig: sneiafit}
\end{figure}

\begin{figure}
\centering \resizebox{12cm}{!}{\includegraphics{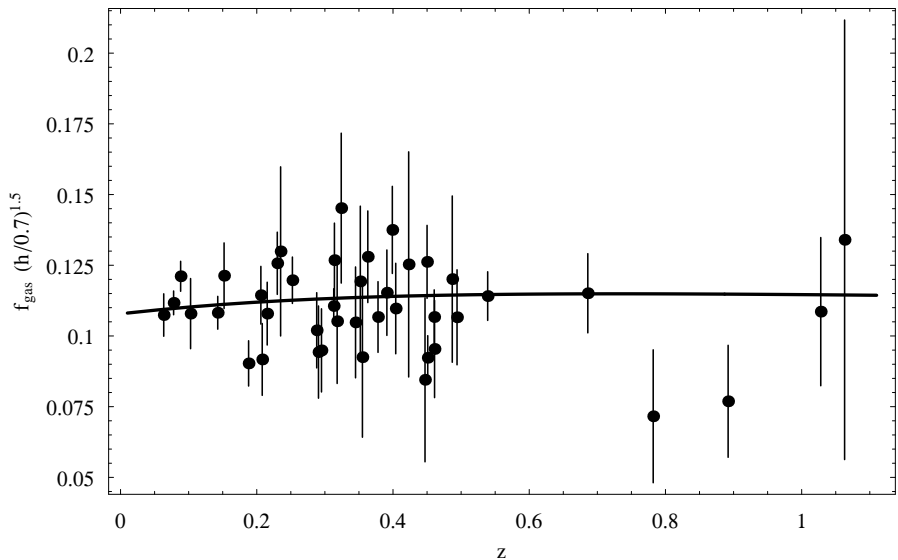}}
\caption{Best fit curve superimposed to the data on the
gas mass fraction. Note that the theoretical curve plots indeed $f_{gas}(z)
{\times} (h/0.7)^{1.5}$ with $h$ set to its best fit value.}
\label{fig: gasfit}
\end{figure}

Best fit model parameters, median values and $1$ and $2 \sigma$ ranges for
the parameters $(\Omega_M, \Omega_{\Lambda}, w_0, h, \Omega_{\phi}$ are
reported in Table 1, while Figs.\,\ref{fig: sneiafit} and \ref{fig: gasfit}
shows how well our best fit model reproduce the data on the SNeIa Hubble
diagram and gas mass fraction.

\begin{table}
\caption{Best fit ($bf$) and median ($med$) values and $1 \sigma$ and $2 \sigma$ ranges of the
parameters $(\Omega_M, \Omega_{\Lambda}, w_0, h,
\Omega_{\phi})$ as obtained from the likelihood analysis.}
\begin{center}
\begin{tabular}{|c|c|c|c|c|}
\hline
Par & $bf$ & $med$ & $1 \sigma$ & $2 \sigma$ \\
\hline \hline
$\Omega_M$ & 0.283 & 0.307 & $(0.272, 0.352)$ & $(0.246, 0.410)$ \\
$\Omega_\Lambda$ & -0.072 & -0.298 & $(-0.54, -0.11)$ & $(-0.92, -0.02)$ \\
$w_0$ & -0.72 & -0.67 & $(-0.74, -0.60)$ & $(-0.79, -0.53)$ \\ $h$ & 0.632
& 0.620 & $(0.588, 0.654)$ & $(0.554, 0.692)$ \\ $\Omega_{\phi}$ & 0.789 &
0.989 & $(0.799, 1.226)$ & $(0.700, 1.574)$ \\
\hline
\end{tabular}
\end{center}
\end{table}

Let us first discuss some general points. First, we note that the best fit
model is in quite good agreement with both the SNeIa and gas data. Indeed,
the $\chi^2$ values are respectively $206$ and $48$ to be contrasted witth
the number of datapoints, being $192$ and $42$ respectively. Less good, but
still in satisfactory agreement with the observed ones, are the values for
the acoustic peak and shift parameters being\,:

\begin{displaymath}
{\cal{A}} = 0.45 \ \ , \ \ {\cal{R}} = 1.67 \ .
\end{displaymath}
Motivated by these results, we can therefore safely conclude that including
a negative $\Lambda$ leads to a model still in agreement with the data so
that this approach to halting eternal acceleration is a viable one from an
observational point of view. In a sense, this is not surprising given that
the net effect of the scalar field and negative $\Lambda$ is to provide a
dark energy fluid with negative pressure dominating the energy budget.
Neveretheless, it is worth stressing that, while the one fluid model has to
resort to a peculiar EoS, the two fluids scenario relies on quite simple
ingredients such as the exponential potential and well motivated
cosmological constant.

Comparing the best fit and median values of each parameter, it is soon
clear that they are well different. This is, however, not a peculiarity of
the model or the analysis, but rather a typical consequence of the non
Gaussian shape of the likelihood. Since the full likelihood ${\cal{L}}({\bf
p}$ is not the product of the marginalized likelihoods
${\cal{L}}_{p_i}(p_i)$, the best fit values ${\bf p}_{bf}$ are not given by
the set of parameters that maximize each single ${\cal{L}}_{p_i}(p_i)$.
Moreover, the marginalized likelihoods are markedly asymmetric so that the
median values may be quite different from the maximum values. This is
particularly the case for $\Omega_{\Lambda}$ because of the constraint
imposed by hand on the negativeness of $\Lambda$. It is worth stressing,
however, that all these caveats are common to the Bayesian approach to
every data fitting problem so that they have not to be considered as model
shortcomings.

A successful fit to the data could, however, be meaningless if the
estimated parameters take unrealistic values. It is therefore worth
discussing how our constraints compare with other previous estimates. Let
us first consider, therefore, the matter density parameter $\Omega_M$. It
is remarkable that the values in Table I are larger than the WMAP3 best fit
model \cite{WMAP3}, being $\Omega_M = 0.268 {\pm} 0.018$. Taken at face values,
the two results are markedly different, even if they well overlap at $1
\sigma$. It is worth stressing, however, that the WMAP3 result reported
above also depends on the cosmological model fitted to the data. For
instance, adding the running of the spectral index to the fiducial
$\Lambda$CDM scenario shifts upward the estimated $\Omega_M$ now giving
$\Omega_M = 0.282 {\pm} 0.020$ in better agreement with our constraint.
Moreover, our estimate also overlaps well with $\Omega_M = 0.27 {\pm} 0.04$
from fitting the Chevallier\,-\,Polarski\,-\,Linder (CPL) model $w(z) = w_0
+ w_1 z/(1 + z)$ \cite{CPL} to a dataset similar to our own \cite{D07}, but
not inlcuding the gas data. As such, we are confident that the higher than
usual matter content of our model is perfectly reasonable and do not
consider this as a worrisome shortcoming. It is worth noting that a higher
value for $\Omega_M$ is somewhat expected. Indeed, in order to compensate
for the attractive effect of the negative cosmological constant, a higher
than usual contribute from the scalar field is needed. But such a large
$\Omega_{\phi}$ should drive the universe towards a kind of
superacceleration that needs to be compensated for by a larger matter
content. The balance among these contrasting effects then induces a pushes
$\Omega_M$ towards larger values. As a final remark, we also note that the
$1$ and $2 \sigma$ ranges are quite larger, while typically $\Omega_M$ is
better constrained in literature. On one hand, this is a consequence of the
degeneracy hinted above among three density parameters $(\Omega_M,
\Omega_{\Lambda}, \Omega_{\phi})$ rather than the only two, $(\Omega_M, \Omega_{DE})$,
typically entering dark energy models. On the other hand, we are not using
here either the CMBR spectrum which probes the very high redshift (where
$\Omega_M$ is the dominant term) or the LSS data (which are sensitive to
the matter content). It is therefore not surprising that the error we
achieve on $\Omega_M$ is so large.

While it is meaningless to compare the value of $\Omega_{\Lambda}$ to the
other estimates in literature given that we are considering a negative
rather than positive $\Lambda$, one could naively think that $w_0$ should
be compared to other determinations. However, it is easy to understand that
this is not the case. Referring to the CPL parametrization as a prototype
for phenomenological EoS and using a similar dataset as our own, Davis et
al. \cite{D07} have obtained $w_0 = -1.1^{+0.4}_{-0.3}$. Compared with the
values in Table 1, we find only a marginal agreement at $2 \sigma$ level.
However, the CPL parametrization assumes a single dark energy fluid, while
our model comprises two different fluids so that the two results are not
comparable. Qualitatively, the smaller is $w_0$ (in absolute value), the
smaller is the driving force speeding up the universe expansion so that one
could be surprised that our model fits so well the data. Actually, what
really matters is not the EoS, but rather than the pressure $p$.
Considering its present day value, we have $p_{\phi}(z = 0) = w_0
\Omega_{\phi} \rho_{crit} = -0.57 \rho_{crit}$ which compares qualitatively
well to the CPL value $p_{CPL} = (1 - \Omega_M) w_0 \rho_{crit} = -0.37
\rho_{crit}$. Moreover, one should also take into account the different $h$ value
which leads to different values of $\rho_{crit}$. Considering that the
difference somewhat fades away when integrating the Hubble parameter to get
the luminosity distance, it is therefore not surprising that the Hubble
diagram is almost the same betweem the one fluid phenomenological CPL model
and our two fluids scenario.

We do not discuss here the constraints on $h$ reported in Table 1 since, as
we have yet said above, this quantity is degenerate with the SNeIa absolute
magnitude. Note that $h$ does not enter either in the acoustic peak nor the
shift parameters so that they do not help in breaking this degeneracy. On
the other hand, the gas mass fraction $f_{gas}(z)$ weakly depends on $h$
since it enters the function $A(z)$ defined in (\ref{eq: defaz}) trhough
$D_A(z)$. It is easy to check that $A(z) \propto (h/h_{\Lambda
CDM})^{-\eta}$ so that for $\eta = 0.214$, we get indeed a negligible
effect. With all these caveats in mind, we do not care much about our
findings on $h$ which turns out to be lower than the $h
\simeq 0.72$ preferred by the local measuremnts \cite{Freedman} and CMBR
data \cite{WMAP3}. Motivated by these considerations, we have therefore
marginalized over $h$ in the discussion of the results, although it is
worth stressing that the difference between $h$ and $h_{\Lambda CDM} = 0.7$
has to be taken into account when comparing $h$ dependent quantities
between the two models.

As a byproduct of the MCMC code, we can also estimate some interesting
derived quantities to be contrasted again measurements in order to get some
more tests on the viability of the model. To this aim, given a quantity
$g({\bf p})$, we compute its value for the points belonging to the MCMC and
estimate its median value and $1$ and $2 \sigma$ ranges as described above.
In order to both speed up the computation and reduce the correlation
between neighbouring points, we have {\it thinned} the chains extracting
one over eight points thus ending up with a sample of $\simeq 11000$ values
which is enough to get reliable estimates.

First, we discuss the results for the acoustic peak parameter whose median
value and $1$ and $2 \sigma$ ranges read\,:

\begin{displaymath}
{\cal{A}} = 0.460 \ \ , \ \ {\cal{A}} \in (0.434, 0.490) \ \ , \ \
{\cal{A}} \in (0.407, 0.523) \ \ .
\end{displaymath}
Pending the question about the validity of its measured value in models
other than the $\Lambda$CDM one, we can therefore conclude that ${\cal{A}}$
predicted by our scenario is in excellent agreement with what is inferred
from the correlation function of LRG galaxies. It is nevertheless worth
extending this analysis by computing the power spectrum and hence the
correlation function predicted by our model, although this is outside our
aims here.

As a second interesting quantity, we consider the shift parameter for which
we find\,:

\begin{displaymath}
{\cal{R}} = 1.38 \ \ , \ \ {\cal{R}} \in (1.21, 1.53) \ \ , \ \ {\cal{R}}
\in (1.06, 1.65) \ \ ,
\end{displaymath}
these values being respectively the median and $1$ and $2 \sigma$ ranges. A
comparison with the measured value, ${\cal{R}} = 1.70 {\pm} 0.03$, should make
us conclude that the model is not able to fit this quantity. While this is
somewhat expected on the basis of the equivalent dark energy model
described in the next section, a more subtle issue has to be taken into
account. Looking at Eq.(\ref{eq: defr}) and considering the constraints on
$\Omega_M$ in Table I, it is easy to conclude that the problem with
${\cal{R}}$ originates from $y(z_{LS}) < y_{\Lambda CDM}(z_{LS})$. We have
checked that this is indeed the case comparing $y(z)$ for our best fit
model with the $\Lambda$CDM curve. Actually, $y(z)$ deviates from
$y_{\Lambda CDM}(z)$ more and more as $z$ increases. Although worrisome,
the smaller value of ${\cal{R}}$ should not be considered as a strong
motivation against our model. As explained in \cite{W07,EM07}, ${\cal{R}}$
is an approximation for the acoustic scale $\ell_a$ which is defined as\,:

\begin{equation}
\ell_a = \frac{\pi (1 + z_{LS}) D_A(z_{LS})}{\int_{0}^{1 +
z_{LS}}{c_s da/(a \dot{a})}} \label{eq: la}
\end{equation}
with $c_s$ the sound speed. The shift parameter is obtained by $\ell_a$
using a different normalization and approximating the denominator as
$\Omega_M^{-1/2}$. This is correct only if the dark energy fades away with
$z$ becoming subdominant at the last scattering. While this is true for the
$\Lambda$CDM and standard quintessence models, this is not for our model as
we will see in the next section. One should therefore compute $\ell_a$ for
our case and compare with the estimated $\ell_a = 303.6^{+1.1}_{-1.2}$
\cite{EM07}. Unfortunately, we are unable to perform such a computation
here because the numerical solution for $\phi(z)$ and hence $a(t)$ becomes
seriously unstable for large $z$. As a consequence, the integral in
Eq.(\ref{eq: la}) turns out to be unreliable so that we can not discuss
anymore this point leaving it for further work.

As a final issue, we also consider the age of the universe. Proceeding as
before, we get (in Gyr)\,:

\begin{displaymath}
t_0 = 13.9 \ \ , \ \ t_0 \in (13.0, 14.9) \ \ , \ \ t_0 \in (12.2, 15.9) \
\ ,
\end{displaymath}
with the same meaning as before for the values reported. Compared with $t_0
= 13.75 {\pm} 0.15 \ {\rm Gyr}$ estimated for the concordance $\Lambda$CDM
model from a combined analysis of the three\,-\,year WMAP, SNeIa and BAO
data \cite{WMAP3}, our result is in very good agreement. It is nevertheless
worth stressing that the WMAP3 constraint is strongly model dependent so
that one should better compare with the model independent measurements.
However, Krauss \& Chaboyer \cite{Krauss} have estimated $t_0 =
12.6^{+3.4}_{-2.6} \ {\rm Gyr}$ from globular clusters, while a similar
value, $t_0 = 12.5 {\pm} 3.5 \ {\rm Gyr}$, has been determined from
nucleochronology \cite{Cayrel}. Given the larger error bars, it is not
surprising that our estimate is in very good agreement with these values
too thus giving a further strong evidence in favour of our scenario.

\begin{figure}
\centering \resizebox{12cm}{!}{\includegraphics{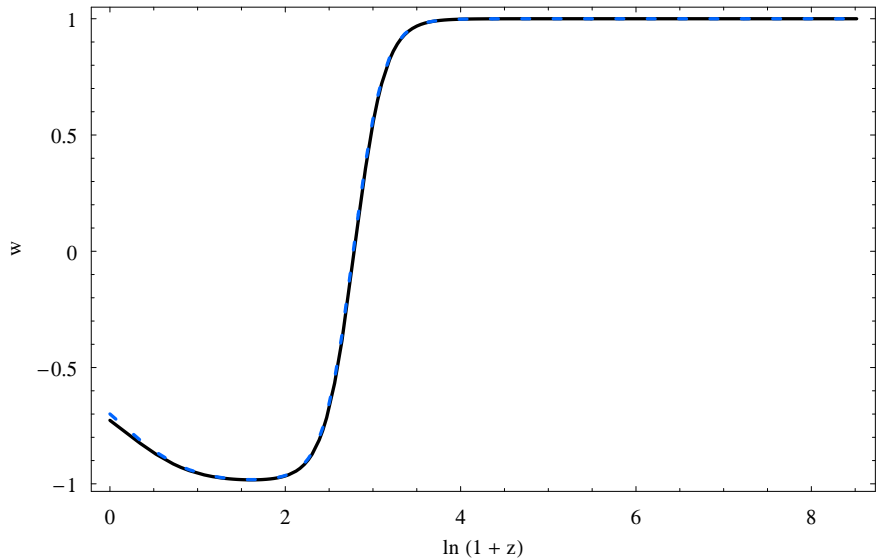}}
\caption{Scalar field (solid line) and effective dark energy
(dashed line) EoS as function of $z$ for the best fit model.}
\label{fig: eosplot}
\end{figure}

\section{Effective dark energy}

The model we are considering is made out of two  {\it non standard} fluids,
namely a negative cosmological constant and a scalar field. From a
phenomenological point of view, however, we can work out a model comprising
a single dark energy fluid giving rise to the same expansion rate as the
present one. Ies EoS may be evaluated as\,:

\begin{equation}
1 + w_{eff}(u) = \frac{(2/3) d\ln{E(u)}/du - \left [ \Omega_M {\rm e}^{3u}
+ \frac{4}{3} \Omega_r {\rm e}^{4u} \right ] E^{-2}(u)}{1 - \left (
\Omega_M {\rm e}^{3u} + \Omega_r {\rm e}^{4u} \right ) E^{-2}(u)}
\label{eq: wdeeff}
\end{equation}
so that the energy density may be then computed as\,:

\begin{equation}
\rho_{DE}(z)/\rho_{crit} = \Omega_{DE}  \exp{\left \{ 3
\int_{0}^{z}{\frac{1 + w_{DE}(z')}{1 + z'} dz'} \right \}} \ .
\label{eq: rhode}
\end{equation}
Fig.\,\ref{fig: eosplot} shows the reconstructed effective dark energy EoS
for our model setting the parameters $(\Omega_M, \Omega_{\Lambda}, w_0)$ to
the best fit value determined by the MCMC code. There are some interesting
issues than can be drawn from this plot. First, we note that $w_{eff}(z)$
and $w_{\phi}(z)$ track each other being essentially
equal\footnote{Actually, in Fig.\,\ref{fig: eosplot}, the two curves for
$w_{\phi}(z)$ and $w_{eff}(z)$ are almost perfectly superimposed so that
they can be hardly distinguished. However, we have checked that the larger
in $\Omega_{\Lambda}$ (in abslute value), the larger is the difference
between the two EoS.} both in the late $(u < 2)$ and early $(u
> 4)$ universe. This is not surprising since, for the best fit values we
are using, the negative cosmological constant is quite small. However, a
similar behaviour is obtained also for other values of the model parameters
so that this is a rather common feature. As such, we will not investigate
in detail how the shape of $w_{eff}(z)$ depends on $(\Omega_M,
\Omega_{\phi}, w_0)$, but our discussion below for the best fit model is
nevertheless quite general.

As it is apparent, $w_{eff}(z)$ cannot be fitted by the most common
phenomenological EoS parametrization such as the CPL one. Indeed, while for
the CPL ansatz $dw/dz = (1 + z)^{-2}$ has a monotonic shape, in our model
$dw_{eff}/dz$ is not at all monotonic and actually also changes its sign.
For the best fit model shown in the plot, $w_{eff}(z)$ first decreases down
to the almost $\Lambda$CDM value $w_{eff} \simeq -0.98$ for $z \simeq 4.0$
and then starts increasing crossing the dust value $w_{eff} = 0$ at $z
\simeq 15$ to finally stays constant to the stiff matter value $w_{eff} =
1$ for $z \ge 210$. Varying $(\Omega_M, \Omega_{\Lambda}, w_0)$ changes
these values, but not the shape of $w_{eff}$ which always asymptotes to 1.
Notwithstanding this unusual feature, the present day values of $w_{eff}$
and $dw_{eff}/dz$ are well within the $2 \sigma$ contours one can obtain
when fitting the CPL formula to the same dataset we have used here. This is
not unexpected given that, for $z < 1$, $D_L(z)$ for a linear EoS closely
matches the luminosity distance\,-\,redshift relation predicted by our
model.

\begin{figure}
\centering \resizebox{12cm}{!}{\includegraphics{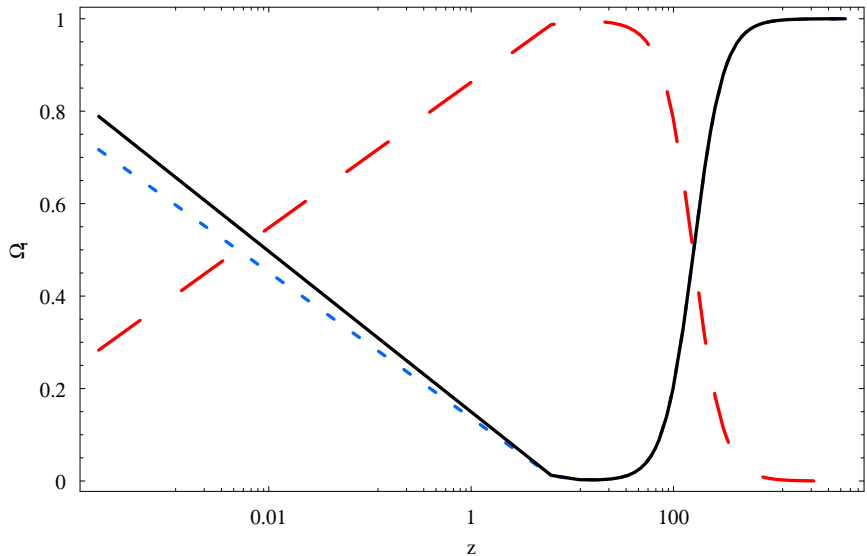}}
\caption{Effective dark energy (black solid), scalar field (blue short dashed)
and matter (red long dashed) density parameters as function of $z$ for the
best fit model.}
\label{fig: omdeplot}
\end{figure}

As an alternative approach, one could resort to a model independent
reconstruction of the EoS (see, e.g., \cite{shaf} and references therein).
Comparing $w_{eff}$ in Fig.\,\ref{fig: eosplot} with Fig. 3 (right panel)
in \cite{shaf} shows us that the effective EoS significantly departs from
what can be reconstructed by the data. It is nevertheless worth noting that
in both cases the EoS asymptotes $w = 1$ even if the transition is still
more rapid in the case of the reconstructed EoS because of the lack of the
initially decreasing phase we have in our model. Although troublesome, we
are confident that the discrepancy between the reconstructed EoS and our
$w_{eff}(z)$ should not be considered as a strong evidence against the
present model. Indeed, data do not tell anything on $w(z)$ directly, but
only on the luminosity (or angular diameter) distance as function of the
redshift. As such, one should not be worry about not reproducing a given
$w(z)$, but the aim of the model must only be to fit the data with
physically meaningful values of the model parameters. This task is
successfully accomplished by our model so that we still retain it as a
valid alternative to a single dark energy fluid scenario.

A potentially more worrisome problem is represented by the behaviour of the
density parameter $\Omega_{DE}(z)$ when compared to the matter one
$\Omega_M(z)$. As it is apparent from Fig.\,\ref{fig: omdeplot},
$\Omega_{DE}(z)$ closely follows $\Omega_{\phi}(z)$ being only slightly
larger because of the need for compensating the negative cosmological
consntat term. From this plot, it is clear that the usual sequence dark
energy\,-\,matter\,-\,radiation dominions is not realized in this model
since the matter era is followed by another dark energy dominated epoch. As
such, the model could be troublesome when dealing with the nucleosynthesis
which takes place during the radiation era. Indeed, it is usually claimed
that, in order the BBN to be efficient, the constraints $\Omega_{DE}({\rm 1
\ Mev}) \le 0.1$ has to be imposed which is strikingly violated by our effective
dark energy model. It is, however, worth noting that such a constraint
usually refers to a negative pressure fluid, while $w_{eff}$ is positive
for our model. Exploring whether BBN can take place in such a background is
outside our aim here, but we will return later on this topic in the
conclusions.

While the BBN epoch is far away in a redshift range where our model could
not be trusted anymore (for reasons explained later), structure formation
typically probes a nearer epoch of the universe evolution. In the usual
scenario, the growth of structure takes place during the matter dominated
era well after the universe has left the radiation epoch. However, in our
case, the radiation dominated era has been replaced by a dark energy
dominated one and, moreover, the matter term is the leading one for a
shorter than usual period. For instance, in the concordance $\Lambda$CDM
scenario, matter drives universe expansion from $z \simeq z_{eq} \sim
3000$, while, for the best fit model in Fig.\,\ref{fig: omdeplot}, the
matter epoch starts later at $z \simeq 157$. Studying in detail the growth
of structure and its impact on observable quantities such as the galaxy
clusters mass function and abundance is outside our aims here and will be
deeply investigated in a forthcoming publication. However, we can
anticipate that this is a subtle issue. On the one hand, denoting with
$\delta$ the matter density fluctuations, one usually has\,:

\begin{equation}
\ddot{\delta} + 2 H \dot{\delta} - 4 \pi G \rho_M = 0
\label{eq: eqdelta}
\end{equation}
so that the dark energy only enters through its effect on the Hubble
parameter $H(z)$. As we have checked, $\Delta H/H = 1 - H/H_{\Lambda CDM}$
decreases from $\simeq 10\%$ to $\simeq 0$ for $z$ ranging between the
beginning ($z \simeq 150$) and the end ($z \simeq 0.45$) of the matter
dominated era, taking $(\Omega_M, \Omega_{\Lambda}, h) = (0.3, 0.7, 0.7)$
for the concordance $\Lambda$CDM model. As such, we do expect that the
growth index $f = d\ln{\delta}/d\ln{a}$ is almost the same as the
$\Lambda$CDM one so guaranteeing a good agreement with this kind of data.
However, Eq.(\ref{eq: eqdelta}) is recovered assuming that dark energy does
not cluster. While this is reasonable for a negative pressure fluid, this
could not be the case for our model since its EoS increases from the
negative value $w_{eff}(z = 0.45) = -0.84$ to $w_{eff}(z = 150) \simeq
0.99$ during the matter era. As such, the dark energy sound speed $c_s^2 =
\partial p_{DE}/\partial \rho_{DE}$ crosses the dust value $c_s = 0$. As such,
it is likely that dark energy can partecipate to the collapse leading to
structure formation thus invalidating Eq.(\ref{eq: eqdelta}) and claiming
for a coupled set of equations.

As a final issue, we would like to comment on a nice feature of the model.
As it is apparent from Fig.\,\ref{fig: omdeplot}, the presence of a dark
energy dominated era before the matter epoch makes it possible to partially
solve the coincidence problem. Indeed, the ratio $r(z) =
\rho_M(z)/\rho_{DE}(z)$ crosses the unit value two times in the universe
history. As a consequence, $r(z)/r_0$ changes by two order of magnitudes
(from 0.1 to 10) over the redshift range $(-0.5, 2.5)$. Converting the
redshift into the dimensionless time $\tau = t/t_0$, we see that $r(z)$
stays {\it close} to its present day value over most of the universe
history thus significantly alleviating the coincidence problem.

\section{Fourth order equivalent theory}

The likelihood analysis performed above has successfully demonstrated that
the model we have introduced in Sect.\,2 is able to reproduce the available
astrophysical data. This encouraging result makes us confident that our
approach to halting the eternal acceleration through the introduction of a
negative cosmological constant is observationally well founded.
Nevertheless, a model comprising two dark ingredients (the scalar field
$\phi$ and the $\Lambda < 0$ term) may be considered as unsatisfactory
because of the need to theoretically motivate two rather than one new
fluids. Moreover, one of them is an unusual negative cosmological constant
which is difficult to reconcile with the classical interpretation in terms
of vacuum energy. As such, it is interesting to work out a possible
reinterpretation of the model in terms of a radically different approach.

Fourth order theories of gravity have recently attracted a lot of attention
as a valid alternative to explain cosmic acceleration without the need of
any dark energy fluid. Moreover, it has been developed a quick method to
find out a $f(R)$ theory giving the same cosmic dynamics of a given dark
energy model. Since all the tests we have considered above rely on the
Hubble parameter, it is obvious that the reconstructed $f(R)$ theory will
fit the data in the same way as the model discussed up to now.

Here, we first summarize the basics of $f(R)$ theories and the method of
reconstructing the gravity Lagrangian from the Hubble parameter $H(z)$ and
then present the application to the model we are considering.

\subsection{Basics of $f(R)$ theories}

Much interest has been recently devoted to the so called curvature
quintessence according to which the universe is filled by pressureless dust
matter only and the acceleration is the result of the modified Friedmann
equations obtained by replacing the Ricci scalar curvature $R$ with a
generic function $f(R)$ in the gravity Lagrangian. The Friedmann equations
therefore read \cite{capozcurv,review}\,:

\begin{equation}
H^2 = \frac{1}{3} \left [ \frac{\rho_m}{f'(R)} + \rho_{curv}
\right ]  \ ,
\label{eq: hfr}
\end{equation}

\begin{equation}
2 \frac{\ddot{a}}{a} + H^2 = - w_{curv} \rho_{curv} \ ,
\label{eq: fried2fr}
\end{equation}
where the prime denotes derivative with respect to $R$, $\rho_{curv}$ is
the energy density of an {\it effective curvature fluid} given as\,:

\begin{equation}
\rho_{curv} = \frac{1}{f'(R)} \left \{ \frac{1}{2} \left [ f(R)  -
R f'(R) \right ] - 3 H \dot{R} f''(R) \right \} \ ,
\label{eq: rhocurv}
\end{equation}
and the barotropic factor of the curvature fluid is\,:

\begin{equation}
w_{curv} = -1 + \frac{\ddot{R} f''(R) + \dot{R} \left [ \dot{R} f'''(R) - H
f''(R) \right ]} {\left [ f(R) - R f'(R) \right ]/2 - 3 H \dot{R} f''(R)}
\label{eq: wcurv}
\end{equation}
Assuming that there is no interaction between matter and curvature terms,
the continuity equation for $\rho_{curv}$ reads \cite{CCT}\,:

\begin{equation}
\dot{\rho}_{curv} + 3 H (1 + w_{curv}) \rho_{curv}  = \frac{3
H_0^2 \Omega_M \dot{R} f''(R)}{\left [ f'(R) \right ]^2}  a^{-3}
\label{eq: curvcons}
\end{equation}
which is identically satisfied as can be easily shown using Eq.(\ref{eq:
hfr}) and expressing the scalar curvature $R$ as function of the Hubble
parameter\,:

\begin{equation}
R = - 6 \left ( \dot{H} + 2 H^2 \right ) \ .
\label{eq: rvsh}
\end{equation}
Combining Eqs.(\ref{eq: hfr}) with Eq.(\ref{eq: fried2fr}) and using the
definition of $H$, one finally gets the following {\it master equation} for
the Hubble parameter \cite{CCT}\,:

\begin{equation}
\dot{H}  -\frac{1}{2 f'(R)} \left \{ 3 H_0^2 \Omega_M a^{-3} + \ddot{R} f''(R) +
 \dot{R} \left [ \dot{R} f'''(R) - H f''(R) \right ] \right \} \ .
\label{eq: presingleeq}
\end{equation}
Inserting Eq.(\ref{eq: rvsh}) into Eq.(\ref{eq: presingleeq}), one ends
with a fourth order nonlinear differential equation for the scale factor
$a(t)$ that cannot be analitically solved also for the simplest cases (for
instance, $f(R) \propto R^n$ unless dust matter contribution is discarded).
Moreover, although technically feasible, a numerical solution of
Eq.(\ref{eq: presingleeq}) is plagued by the large uncertainties on the
boundary conditions (i.e., the present day values of the scale factor and
its derivatives up to the third order) that have to be set to find out
$a(t)$ by solving Eq.(\ref{eq: presingleeq}).

Given these mathematical difficulties, a different approach has been
proposed in \cite{CCT} (hereafter CCT) where Eq.(\ref{eq: presingleeq}) is
considered as a way to determine $f(R)$ rather than $a(t)$. Rearranging the
different terms suitably, CCT obtained a linear third order differential
equation for $f$ in terms of the redshift $z = 1/a - 1$ (having set $a_0 =
1$) that can be easily solved numerically for a given $H(z)$. By this
method, it is then possible to find out which $f(R)$ theory reproduces the
same dynamics of a given dark energy model, thus showing a formal
equivalence between these two radically different approaches.

CCT developed the method using the redshift $z$ as integration variable
since it is common to have an analytical expression for the Hubble
parameter as function of $z$. However, this is not the case for the model
considered here, so that it turns out to be numerically best suited to use
$t$ as integration variable. To this aim, we follow \cite{ExpPot} to
rewrite the main formulae in CCT to finally get the equation determining
$f(t)$ is then\footnote{With an abuse of notation, we write $f(t)$ rather
than $f[R(t)]$.}\,:

\begin{equation}
{\cal{H}}_3(t) \frac{d^3f}{dt^3} + {\cal{H}}_2(t)
\frac{d^2f}{dt^2} + {\cal{H}}_1(t) \frac{df}{dt} = - 3 H_0^2
\Omega_M \dot{R}^2 a^{-3}(t) \ , \label{eq: singleeq}
\end{equation}
with\,:

\begin{equation}
{\cal{H}}_1 = 2 \dot{H} \dot{R} + H \ddot{R} + 2 \ddot{R}^2
\dot{R}^{-1} - d^3R/dt^3 \ ,
\end{equation}

\begin{equation}
{\cal{H}}_2 = - \left ( 2 \ddot{R} + H \dot{R} \right ) \ ,
\end{equation}

\begin{equation}
{\cal{H}}_3 = \dot{R} \ .
\end{equation}
where $R$ is given by Eq.(\ref{eq: rvsh}). In order to integrate
Eq.(\ref{eq: singleeq}), we need to specify boundary conditions
that are more conveniently assigned at the present time. We
slightly generalize here the discussion presented in CCT. First,
let us remember that, in a fourth order theory, we may define an
{\it effective gravitational constant} as $G_{eff} = G_N/f'(R)$,
with $G_N$ the usual Newtonian gravitational constant. Its rate of
variation will be given as\,:

\begin{equation}
\frac{\dot{G}_{eff}}{G_{eff}} = - \frac{1}{t_0}
\frac{f''(R)}{f'(R)} \frac{dR}{d\tau} \ . \label{eq: geffvar}
\end{equation}
It is quite natural to assume that the effective and the Newtonian
gravitational couplings take the same values today, so that we get
the condition\,:

\begin{equation}
f'(R_0) = 1 \ . \label{eq: in1}
\end{equation}
Evaluating Eq.(\ref{eq: geffvar}) for $t = t_0$ (i.e., $\tau =
1$), we may determine $f''(R_0)$ provided an estimate of
$(\dot{G}_{eff}/G_{eff})_{t= t_0}$ is given. Since in our theory
$G_N$ is constant, we may assume that the measurements of the
variation of $G_N$ \cite{uzan} actually refers to $G_{eff}$ and
use these results to get an estimate of
$(\dot{G}_{eff}/G_{eff})_{t= t_0}$. We thus take as our boundary
condition\,:

\begin{equation}
f''(R_0) = - t_0 \left ( \frac{\dot{G}_{eff}}{G_{eff}} \right
)_{obs} \left ( \frac{dR}{d\tau} \right )^{-1} \ , \label{eq: in2}
\end{equation}
having used Eq.(\ref{eq: in1}). Finally, inserting Eqs.(\ref{eq:
in1}) and (\ref{eq: in2}) into Eq.(\ref{eq: rhocurv}) and then in
(\ref{eq: hfr}) evaluated today, we get\,:

\begin{eqnarray}
f(R_0) & = & 6 H_0^2 \left ( 1 - \Omega_M + \frac{R_0}{6 H_0^2}
\right ) f'(R_0) \nonumber \\ ~ & + & 6 H_0 \left ( \frac{dR}{dt}
\right )_{t = t_0} f''(R_0) \ . \label{eq: in0}
\end{eqnarray}
From Eqs.(\ref{eq: in2})\,-\,(\ref{eq: in0}), the following boudary
conditions straightforwardly descend\,:

\begin{equation}
\left ( \frac{df}{dt} \right )_{t = t_0} = \left ( \frac{dR}{dt}
\right )_{t = t_0} f'(R_0) \ , \label{eq: in1bis}
\end{equation}

\begin{equation}
\left ( \frac{d^2f}{dt^2} \right )_{t = t_0} = \left (
\frac{dR}{dt} \right )_{t = t_0}^2 f''(R_0) + \left (
\frac{d^2R}{dt^2} \right )_{t = t_0} f'(R_0) \label{eq: in2bis}
\end{equation}
that have to be used, together with Eq.(\ref{eq: in0}), to
numerically solve Eq.(\ref{eq: singleeq}). Combining the solution
thus obtained for $f(t)$ with $R(t)$ evaluated through
Eq.(\ref{eq: rvsh}), one finally finds $f(R)$ thus recovering the
higher order theory that mimicks the assigned dark energy model.
We refer the reader to \cite{CCT,ExpPot,danish} for some
interesting examples.

A preliminary comment is in order here, however. The dark energy model and
its $f(R)$ duality representation share the same expression for the Hubble
parameter $H(z)$ (and hence for the scale factor dependence on cosmic
time). As such, all the observational tests relying directly on $H(z)$ (as
the acoustic peak parameter ${\cal{A}}$) or its integral (as the luminosity
distance or the age of the universe) are not able to discriminate between a
whatever dark energy model and its $f(R)$ counterpart. Nevertheless, such
an equivalence only holds at the background level. Indeed, since the
underlying theory of gravity is radically different, at the perturbative
level, the two models make clearly distinct predictions. As a consequence,
observational tests relying on the solution of perturbation equations (such
as the growth factor and the power spectrum) could be able to break this
degeneracy. Actually, at the moment, observational determinations of the
growth index $d\ln{\delta}/d\ln{a}$ are still quite noisy, while the
uncertainties in the relation betwenn the observationally derived galaxy
power spectrum and the theoretically predicted matter one still prevent
from discriminating among dark energy and modified gravity. Moreover, we
only rely here on tests depending on $H(z)$ so that the $f(R)$ theory we
will reconstruct is observationally equivalent to our two fluids scenario.

\subsection{Reconstruction of $f(R)$}

All we need to apply the procedure described above is an expression for the
Hubble parameter $H$ as function of the cosmic time $t$. For the model we
are considering, this may be obtained numerically as described in Sect.\,2
so that we only have to choose a value for $\dot{G}_{eff}/G_{eff}$ in order
to set the initial condition for $f''(R_0)$. We fix $\dot{G}_{eff}/G_{eff}
= 0$ in good agreement with most of the estimates in \cite{uzan}. We have also
checked that changing this value within the quoted uncertainties
does not affect significantly the main results.

Before presenting the reconstructed gravity Lagrangian, there is a
conceptual point to clarify. The method described above represents a sort
of {\it bridge} between two different scenarios. In particular, their
matter contents could be different, so that we should define both
$\Omega_M^{curv}$ and $\Omega_M^{DE}$ to denote this quantity in the two
different models. In principle, there is no reason why $\Omega_M^{curv} =
\Omega_M^{DE}$ should hold. However, since $\Omega_M^{DE}$ is close to the
fiducial value ($\Omega_M \simeq 0.3$) suggested by model independent
estimates (e.g., from galaxy clusters abundance), we take $\Omega_M^{curv}
= \Omega_M^{DE}$. Should we have chosen a different value for $\Omega_M^{curv}$,
we have had a different reconstructed $f(R)$, but still providing the same
Hubble parameter as the model with the scalar field and the negative
$\Lambda$ we have tested against data. To be more precise, one could say
that our method is not able to recover a single $f(R)$, but rather a class
of $f(R)$ models parameterized by $\Omega_M^{curv}$. An astrophysics based
estimate of $\Omega_M$ is the only way to break this degeneracy making it
possible to select the most suitable member in this family of fourth order
theories.

\begin{figure}
\centering \resizebox{12cm}{!}{\includegraphics{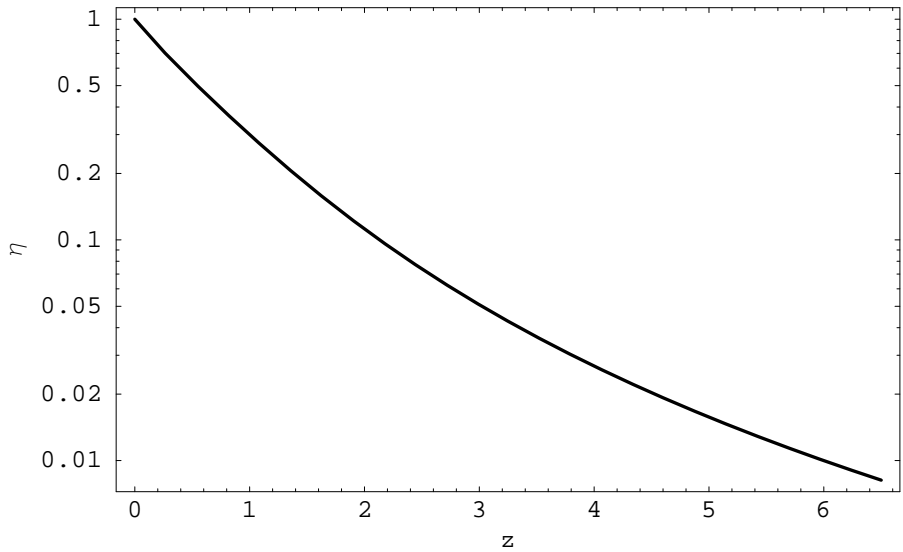}}
\caption{$\eta$ as function of the redshift $z$ for the $f(R)$
reconstructed from the best fit model. Here we define $\eta =
\left [ f(R)/R \right ]/\left [ f(R)/R \right ]_{z = 0}$, see the
text for further details.}
\label{fig: etavsz}
\end{figure}

With this remark in mind, we have performed the reconstruction of $f(R)$
for the model presented in Sect.\,2 setting its parameters $(\Omega_M,
\Omega_{\Lambda}, w_0)$ to their best fit values. Rather than reconstructing $f(R)$
directly, it is more instructive to first consider\,:

\begin{equation}
\eta = \left [ \frac{f(R)}{R} \right ] {\times} \left [
\frac{f(R)}{R} \right ]_{z = 0}^{-1} \label{eq: defeta}
\end{equation}
which is identically 1 for the Einstein\,-\,Hilbert Lagrangian, $f(R) = R$.
Departures of $\eta$ from unity therefore quantifies how much the modified
Lagrangian should depart from the standard one in order to fit the data as
well as the input Hubble parameter.

\begin{figure}
\centering \resizebox{12cm}{!}{\includegraphics{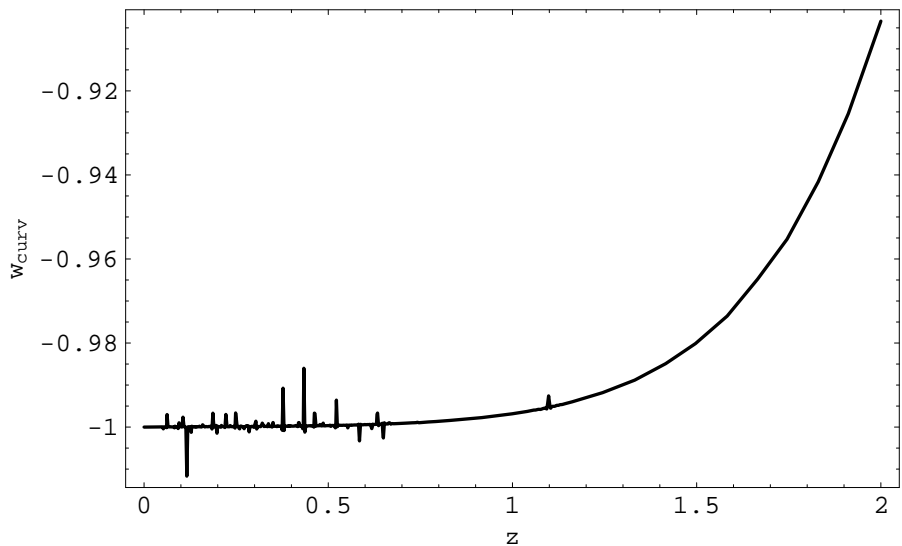}}
\caption{The EoS $w_{curv}$ of the effective curvature fluid for the reconstructed $f(R)$
theory. Note that the small fluctuations are only artefacts due to
numerical noise.}
\label{fig: wcplot}
\end{figure}

To this aim, it is interesting to look at Fig.\,\ref{fig: etavsz} where we
report $\eta$ as function of $z$ over the redshift range $0 \le z \le 6$,
extending well above the one probed by the data we have used. It is
apparent that $\eta$ quickly departs from 1, i.e. the reconstructed $f(R)$
dramatically differs from the Einsteinian one. This is not unexpected at
all. Indeed, in order to fit the data with a matter only universe, one
should strongly modifies the gravity Lagrangian at least for values of $R_s
= R/R_0$ corresponding to the accelerating expansion era. For
larger values of $z$, the universe is decelerating so that one could
naively expect to recover a standard situation with $\eta$ approaching $1$.
This is clearly not the case. Actually, this somewhat counterintuitive
result is easily explained reconsidering Fig.\,\ref{fig: omdeplot} which
shows that in our model dark energy dominates again after a finite period
of matter domination. In order to recover this unusual behaviour, $f(R)$
must still depart from the General Relativity Lagrangian.

Although the reconstructed $f(R)$ is obtained numerically, it is useful to
have an analytical approximation which we indeed find to be given by\,:

\begin{equation}
f(R) = f_0 R_s \left [ 1 + \left ( \alpha R_s^n + \beta R_s^{-m} \right )
\ln{R_s} \right ]
\label{eq: approx}
\end{equation}
with $R_s = R/R_0$, $(\alpha, \beta, n, m)$ real parameters to be fitted to
the numerical data. We have checked that such a formula reproduces the
reconstructed $f(R)$ within less than $4\%$ over the full redshift range we
have probed. For the best fit model, we get\,:

\begin{displaymath}
(\alpha, \beta, n, m) = (-0.38, -0.61, -0.15, 0.70) \ .
\end{displaymath}
Changing the values of $(\Omega_M, \Omega_{\Lambda}, w_0)$ alters these
values, but Eq.(\ref{eq: approx}) still provides a very good approximation.
We stress, however, that the approximating formula (\ref{eq: approx}) has
been tested only over the redshift range $(0, 6)$ so that cannot be
extrapolated to larger values. As such, one has not to give any weight to
the fact that $f(R)$ does not reduce to $R$ in the very early universe as
expected to recover the BBN succesfull results.

It is interesting to consider the EoS $w_{curv}$ of the effective curvature
fluid which is plotted in Fig.\,\ref{fig: wcplot}. This plot shows that the
curvature fluid actually behaves as a sort of usual cosmological constant
with its EoS increasing with resdshift, but still remaining quite close to
the $\Lambda$ value $w = -1$ over the full redshift range probed by the
data. This result could be somewhat anticipated noting that the
reconstructed $f(R)$ theory provide the same $H(z)$ and hence the same
$\Delta H/H$ as the starting model. Since this quantity stays within $10\%$
over the range $0 \le z \le 6$, it is not surprising that the effective
curvature fluid behaves as an almost $\Lambda$ term. However, we have
checked that $w_{curv}$ departs more and more from $w = -1$ as $z$
increases even if we are not able to draw definitive conclusions because of
the dramatical increase of the numerical noise.

As a final comment, it is worth stressing again that the reconstructed
$f(R)$ provides the same cosmic dynamics (i.e., scale factor and Hubble
parameter) of a model comprising both a quintessential scalar field and a
negative cosmological constant. As a consequence, being the background
dynamics exactly the same, the $f(R)$ model based on Eq.(\ref{eq: approx})
predicts the same behaviour for the luminosity distance as function of the
redshift $z$ (and hence the same SNeIa Hubble diagram and gas mass fraction
vs $z$ curve) and equal numerical values for the acoustic peak and shift
parameters. As such, all the observational tests in Sect. 3 successfully
met by the negative $\Lambda$ model are equally satisfied by its $f(R)$
duality representation constructed here. Such a fourth order theory is
therefore able to both fit the available data and avoid eternal
acceleration. In particular, the net effect is to mimic an unusual negative
cosmological constant with a modified gravity Lagrangian and no dark
energy. This interesting result shows that a $\Lambda < 0$ term could
result as a net effect of forcing the gravity theory to be an Einsteinian
rather than a fourth order one. Moreover, halting eternal acceleration is
therefore possible without adding any {\it ad hoc} ingredient to the cosmic
pie.

\section{Conclusions}

The unprecedented high quality data accumulated in the recent years have
depicted the scenario of a spatially flat universe presently undergoing a
phase of accelerated expansion thus motivating the lot of interest devoted
to the search of viable candidates to drive this cosmic acceleration.
Although many possible models have been put on the ground, almost all of
them predict a two phase scenario with standard dust matter dominating the
first epoch of decelerated expansion and a (actual or effective) dark
energy fluid fueling the present day cosmic speed up. Notwithstanding the
underlying mechanism, the fate of the universe seems to be yet written. As
matter fades away, dark energy becomes more and more dominant leading to an
eternal acceleration\footnote{Here, we are not considering phantom dark
energy which leads to a superaccelerated expansion ending in a Big Rip like
singularity. Note, however, that phantom scenarios are affected by the same
problems of eternal acceleration till the cosmic doomsday.} so that serious
problems arise with the formulation of the ${\cal{S}}$ matrix and quantum
field theory.

In an attempt to avoid these problems, we have explored here a model made
out of three rather than two ingredients. Besides the usual dust matter and
radiation and a scalar field responsible of cosmic acceleration, a negative
cosmological constant has been added to the energy density budget. While
being subdominant during the matter dominated decelerated phase and the
scalar field dominated accelerated epoch, this new component may contrast
the action of the quintessential scalar field during the future evolution
of the universe. Depending on the balance between these two terms, eternal
acceleration may be avoided leading to a Big Crunch collapse in a finite
time. This possibility renders the model quite attractive from a
theoretical point of view and have motivated us to further explore its
viability through a comparison with the available astrophysical data. To
this aim, we have fitted the model to the most recent release of SNeIa data
(made out of data coming from the SNLS, ESSENCE and GOODS surveys) and the
gas mass fraction in galaxy clusters also setting a prior on the acoustic
peak and shift parameters. This large dataset has been successfully
reproduced thus showing that the presence of a negative $\Lambda$ is fully
consistent with the data. As such, our model may be considered as a viable
solution to the problem of eternal acceleration.

It is, however, worth investigating whether the same model may be
interpreted as an effective representation of a radically different
scenario. Such a possibility is offered by fourth order theories of gravity
according to which cosmic acceleration is the result of a modified gravity
Lagrangian with the scalar curvature $R$ replaced by an analytic function
$f(R)$ to be determined by the data. Using a method developed by one of us
\cite{CCT,ExpPot}, we have reconstructed $f(R)$ by imposing that the cosmic
dynamics is the same as the one determined by the action of the scalar
field and the negative $\Lambda$. Since the Hubble parameter is the same in
both scenarios, both models are able to fit the data in the same way.
Moreover, the reconstructed $f(R)$ does not include any cosmological
constant term thus showing that a negative $\Lambda$ could also be the
consequence of forcing a fourth order Lagrangian to the linear Einsteinian
one.

Although encouraging, these results must be considered only preliminary.
Indeed, the model is affected by some undesirable shortcomings when applied
to the early universe. First, the model turns out to be matter dominated
only for a finite redshift range, while its expansion is scalar field
driven during most of the high redshift range. Moreover, the scalar field
EoS converges towards the stiff matter value $w = 1$ so that one could
worry about the growth of structure and the CMBR anisotropy spectrum. Both
these results are not fully unexpected. Indeed, it has yet been
demonstrated that an exponential potential may lead towards a scalar field
dominated universe. Our result shows that this conclusion also holds in the
case a negative $\Lambda$ is present as could be predicted in advance
noting that this term becomes quickly subdominant at large $z$ with respect
to the matter and the scalar field ones. One could therefore consider the
possibility to replace the exponential potential with a different $V(\phi)$
using, e.g., the sum of two exponential with opposite signs. Suitably
weighting the two terms may lead to a model which is similar to the present
one at small $z$ thus giving the same successful match with observations.
On the other hand, making the additional term dominant for large $z$ may
solve the problems of the present model in the early universe. We stress,
however, that here we have been more interested in looking for a model able
to halt eternal acceleration. We have therefore paid more attention to the
recent and future universe so that some problems with the early universe
were foreseeable. The complicated task of matching the early, late and
future expansion looking for a suitable potential will be addressed in a
forthcoming work.

Further developments of the approach pursued here are possible from the
observational and the theoretical point of view. On one hand, we have only
tested the dynamics of the model, but severe constraints may be obtained
investigating structure formation (see, e.g., \cite{exppert} for a
discussion of this issue for the expoential potential model). To this aim,
a first important step may be done solving the perturbation equation in the
linear regime thus determining the growth factor $D = \delta/a$ and the
growth index $f = d\ln{\delta}/d\ln{a}$. Since a negative $\Lambda$ has the
same effect of an attractive gravitational potential, it is possible that
the collapse of structures is enhanced so that it is worth investigating
what constraints may be imposed by requiring that this effect does not
change significantly neither $D$ nor $f$ with respect to the successful
$\Lambda$CDM predictions. It is worth stressing, at this point, that which
are the perturbation equations to solve depend on the scenario adopted.
Should we interpret the model in terms of $f(R)$ theories, the standard
theory of linear perturbations must be abandoned and different equations
have to used. This gives us the possibility to select between the two
equivalent descriptions retaining the one which predicts $D(a)$ and $f(a)$
most similar to the $\Lambda$CDM ones since they are more likely able to
fit the large scale structure data.

On a different ground, one can arguably discriminate between the two
components dark energy model and its $f(R)$ duality representation based on
Solar System tests. Indeed, in the low energy limit, the static and
spherically symmetric solution of the Einstein equations for the two
components dark energy model leads to the usual Newtonian potential (apart
some negligible deviations introduced by the tiny neative $\Lambda$ and the
scalar field) so that all the classical tests of gravity are automatically
verified. On the other hand, the low energy limit of $f(R)$ theories may
lead to significant deviations from the Newtonian potential. As such, these
theories are indeed severely constrained by the Solar System tests so that
some models have been carefully designed just to evade this problem
\cite{viablefr}. Although a strong debate is still present on which is the
the correct way to work out the low energy limit of such theories, it is
worth investigating this issue with some detail in a future work in order
to see whether our approximated $f(R)$ expression (\ref{eq: approx}) is
indeed viable from this point of view. Should this not be the case, we
could give it away, but still retaining the two components dark energy
model as a viable way to halt eternal acceleration.

A more {\it philosophical} comment is in order. Along the work presented
here, we have first introduced a two components dark energy model
comprising a negative $\Lambda$ and an exponential potential scalar field
$\phi$. After successfully testing it against the data, we have worked out
its fourth order counterpart determining the function $f(R)$ entering the
modified gravity Lagrangian giving the same cosmic expansion $H(z)$. A
guidance idea has been the Occam's razor that suggests to give out models
with an unnecessary large number of unknown elements. The price to pay has
been to replace the simple general relativistic $f(R) = R$ with a fourth
order expression analytically approximated as in Eq.(\ref{eq: approx}). One
could wonder whether such a complicated Lagrangian may indeed be in line
with the Occam's razor spirit. Actually, we are at the moment unable to
find out a fundamental theory leading to our reconstructed $f(R)$, although
some Lagrangian containing the sum of terms with different powers of $R$
have been studied in literature (see, e.g., third reference in
\cite{MetricRn}). Should a fundamental theory not be found, one can indeed
consider {\it unnecessary} our modification of $f(R)$ and put our theory on
the same ground as the two components dark energy model we have started
from.

As a conclusive remark, we believe it is worth stressing that halting
eternal acceleration is possible. Using a two component model comprising a
quintessential scalar field and a negative $\Lambda$ represents the easiest
way to generate a transient cosmic acceleration. Fourth order theories make
it possible to interpret both these (somewhat problematic) terms as
effective manifestations of a different scenario. The reconstructed $f(R)$
may then be seen as the missing bridge between the easy world of many
components models and the mathematically complicated structure of higher
order gravity theories. \\

\ack VFC is grateful to S. Capozziello and A. Troisi for
the interesting discussions on the manuscript and to V. Salzano for help
with the MCMC programming.

\end{document}